\documentclass[12pt]{article}
\usepackage{latexsym,amssymb,bbm,newlfont}
\usepackage{natbib,endnotes}
\bibpunct[,~]{(}{)}{;}{a}{}{,~}
\def\aycite{\cite}
\def\ben{\begin{enumerate}}
\def\een{\end{enumerate}}
\def\bi{\begin{itemize}}
\def\ei{\end{itemize}}
\def\bd{\begin{description}}
\def\ed{\end{description}}
\def\bq{\begin{quote}}
\def\eq{\end{quote}}
\def\bc{\begin{center}}
\def\ec{\end{center}}
\def\be{\begin{equation}}
\def\ee{\end{equation}}
\def\bea{\begin{eqnarray}}
\def\eea{\end{eqnarray}}

\def\f{\varphi}   
\def\g{\gamma}

\def\l{\lambda}
\def\m{\mu}
\def\n{\nu}

\def\q{\theta}

\def\del{\partial}
\def\ket#1{ | #1 \rangle }

\begin{document}
\sloppy

\thispagestyle{empty}

\vspace*{-8mm}

\renewcommand{\thefootnote}{\fnsymbol{footnote}}

\noindent{Penultimate version, forthcoming in
\emph{International Studies in the Philosophy of Science}.
June 2008.%
\footnote{Earlier versions of this paper had been presented at
the ESF Conference ``Philosophical and Foundational Problems in Quantum Theory'' 2005 in Budapest
and the DPG Spring Meeting 2006 in Dortmund.
}

\vspace*{18mm}

\noindent{\Huge Does the Higgs Mechanism Exist?}

\vspace*{5mm}

\noindent{\large Holger Lyre}%
\footnote{Philosophy Department, University of Bonn, E-mail: lyre@uni-bonn.de}

\vspace*{15mm}

\noindent{\em
This paper explores the argument structure of the concept
of spontaneous symmetry breaking in the electroweak gauge theory
of the Standard Model: the so-called Higgs mechanism.
As commonly understood, the Higgs argument is designed
to introduce the masses of the gauge bosons by a spontaneous breaking
of the gauge symmetry of an additional field, the Higgs field.
The technical derivation of the Higgs mechanism, however,
consists in a mere re-shuffling of degrees of freedom
by transforming the Higgs Lagrangian in a gauge-invariant manner.
This already raises serious doubts about the adequacy of the
entire manoeuvre. It will be shown that no straightforward
ontic interpretation of the Higgs mechanism is tenable
since gauge transformations possess no real instantiations.
In addition, the explanatory value of the Higgs argument
will be critically examined.}

\vspace{15mm}


\section{A Short Historical Introduction}

In 1961 Sheldon Glashow presented the first $SU(2) \times U(1)$ gauge theoretic model
for the electroweak interaction. The model was based on a straightforward application
of the gauge principle: the idea that in order to fulfil the requirement of
invariance of the fundamental Lagrangian under local gauge transformations
(of the considered symmetry), one needs an inhomogeneous coupling term.
Mathematically, the usual derivative is to be replaced by an appropriate gauge
covariant derivative ($\partial_\mu \to \partial_\mu + i q A_\mu$ for U(1) for instance).%
\endnote{\label{recipe}Of course this well-known recipe does not entail
the existence of a non-zero interaction field (compare footnote \ref{gaugeprin}).
The logic of the gauge argument has been unfolded by several authors
in recent years; cf.\
\citet{brown99}, \citet{lyre2001} and \citet{martin2002}.}
Yet, the gauge potential term that thus occurs does not include a mass
and so a straightforward application of the gauge principle leads
to massless gauge bosons only. Since the weak interaction has short range,
it was clear that the gauge bosons were required to be massive.
At that time, this could have been seen as a sign that demanding
gauge invariance is to proceed on the wrong track, and that the idea of extending
the gauge principle to symmetries higher than U(1)---discovered only half a decade
earlier by Yang and Mills---is not fundamental. On the other hand it was also known
that gauge symmetry was seemingly vital for the construction of renormalizable
quantum field theories (QFTs). There was thus a tension between the requirement
of massive exchange particles for the weak interaction and the renormalizability
of an appropriate QFT, a tension that Glashow could not dissolve within his early work.

By 1964, Peter Higgs and also Brout, Englert, and Kibble extended
the work of Goldstone on spontaneous symmetry breaking to gauge theories.
But they did not apply their framework to any phenomenologically relevant model.
This was first done by Abdus Salam and Steven Weinberg in 1967 and 1968
for Glashow's $SU(2) \times U(1)$ electroweak model.
The key idea was that $SU(2) \times U(1)$ gauge symmetry is an exact,
but ``hidden'' symmetry, and that masses can be generated ``dynamically''
by spontaneous symmetry breaking. Salam and Weinberg succeeded in completing
a framework which indeed dissolves the above-mentioned tension between massive
exchange particles and renormalizability.
This framework is the well-known Standard Model's
\emph{GSW electroweak theory} based on $SU(2) \times U(1)$ gauge symmetry---its
first impressive experimental confirmation was the discovery
of weak neutral currents in 1973 at CERN.
Thus, the GSW theory implements the existence of massive
exchange particles, the weak W- and Z-gauge bosons, as well as massive
leptons by means of the now widely-known ``Higgs mechanism'':
the ``spontaneous breaking of a gauge symmetry.''

The enormous importance of GSW can perhaps be measured by the succession
of Nobel prizes ``induced'' by it.
In 1979 the prize was awarded to ``GSW'' themselves (Glashow, Salam, Weinberg);
next, after the 1983 discovery of W- and Z-bosons at CERN,
Carlo Rubbia and Simon van der Meer were awarded the 1984
Nobel prize for their leadership of the experiments;
and finally Gerard 't Hooft and Martinus Veltman were honoured in 1999 for
their mathematical proof in the early 1970s, by which the renormalizability
of spontaneously broken gauge theories was theoretically established.

Parts of the 1979 Nobel lectures show the importance of the idea
of the Higgs mechanism within the GSW theory and its conceptual understanding
not only by the common physicist but also by the leading figures.
Glashow, for instance, writes:
\bq
In pursuit of renormalizability, I had worked diligently
but I completely missed the boat. The gauge symmetry is an exact symmetry,
but it is hidden. One must not put in mass terms by hand.
The key to the problem is the idea of spontaneous symmetry breakdown ...
Salam and Weinberg ... first used the key. \cite[498]{lundqvist92}
\eq
And Weinberg continues along the same lines:
\bq
Higgs, Kibble, and others ... showed that if the broken
symmetry is a local, gauge symmetry, like electromagnetic gauge invariance,
then although the Goldstone bosons exist formally, and are in
some sense real, they can be eliminated by a gauge transformation, so that
they do not appear as physical particles. The missing Goldstone bosons
appear instead as helicity zero states of the vector particles, which thereby
acquire a mass. \cite[545]{lundqvist92}
\eq

We will see in a moment how the detailed derivation of
the Higgs mechanism works.
To be sure, there is nothing wrong with the mathematics of it,
but on closer inspection of the ``mechanism'' it will become clear
that a deeper conceptual understanding of the formalism is not at all
as obvious and as straightforward as most presentations, notably textbook
presentations, of the Higgs mechanism usually pretend.
For instance---and as the alert philosophy of physics reader will certainly
have noticed already---the status of the symmetries in question,
gauge symmetries, is in fact a non-empirical or merely conventional one
precisely in the sense that neither global nor local gauge transformations
possess any real instantiations (i.e.\ realizations in the world).
Rather their status is comparable to the status of coordinate transformations
(the status of gauge symmetries will be addressed in detail in Sec.~\ref{ontol}).
How is it then possible to instantiate a mechanism, let alone a dynamics
of mass generation, in the breaking of such a kind of symmetry?
Suspicions like this should raise philosophical worries about
the true ontological and explanatory story behind the Higgs mechanism.

Although in recent times interest in spontaneous symmetry breaking (SSB)
within philosophy of physics has risen, no one has yet scrutinized
the Standard Model Higgs mechanism in the particular direction just indicated.
Some authors, e.g.\ \cite{castellani2003}, \cite{kosso2000b} and
\cite{morrison2003}, are interested in the epistemological status of a ``hidden''
symmetry and in whether and how one is justified in building physical models
on unobserved symmetries. While this is certainly an interesting topic,
these authors nevertheless consider SSB scenarios in various branches of physics
mainly on a par---and this, as we will see, is a serious misunderstanding.
Chuang Liu in a series of papers carefully analyzes and compares
the various sorts of SSBs that play a role
in classical physics \citep{liu2003b}, quantum statistics \citep{liu+emch2005},
and condensed matter as well as particle physics \citep{liu2002b}.
In this latter paper Liu correctly emphasizes some important disanalogies
between the concept of SSB in the well-known ferromagnet model on the one hand
and in particle physics on the other hand, which we will also discuss,
but he unfortunately does not delve into the important question
of the meaning of breaking a conventional gauge symmetry.
John Earman (\citeyear{earman2003b}, \citeyear{earman2004a})
perhaps comes closest to our particular suspicion
when he writes:
\bq
As the semi-popular presentations put it,
``Particles get their masses by eating the Higgs field.''
Readers of \emph{Scientific American} can be satisfied with these just-so stories.
But philosophers of science should not be. For a genuine property like mass cannot be
gained by eating descriptive fluff, which is just what gauge is. Philosophers of science
should be asking the Nozick question: What is the objective (i.e.\ gauge invariant)
structure of the world corresponding to the gauge theory presented in the
Higgs mechanism? \cite[1239]{earman2004a}
\eq
Indeed, how can any physical mechanism arise from the breaking of
a merely conventional symmetry requirement?
(Similarly, one would not think that any physics flows out of
the breaking of coordinate invariance!---Again this will be
addressed in detail in Sec.~\ref{ontol}.)
Earman himself unfortunately only touches on the issue without
really answering it.%
\endnote{A reaction on Earman certainly worth reading and tentatively
in spirit with the present paper, though not as decisive as we are
concerning ontological consequences, is \cite{smeenk2006}.
Chris Smeenk and I wrote our papers independently and only discovered
certain similarities in our views only after a recent meeting at a conference.}
We are in fact left here with a series of pressing questions still in the air.
In what sense, we may ask, are Goldstone bosons ``real'' (\`a la Weinberg)?
In what sense are the masses of the particles truly ``dynamically generated''?
What exactly is the predictive and explanatory power of the Higgs mechanism?
And, finally, does this very mechanism ``exist'' at all?


\section{The Higgs Mechanism}


\subsection{Ferromagnetism as a Case of SSB}
\label{ferromagnet}

Before we delve into the Higgs mechanism, it will be instructive
to take a look at ferromagnetism first as \emph{the} paradigm case of SSB.
This is all the more useful since almost any presentation of the Higgs mechanism
stresses the supposed analogy between the Higgs case and the ferromagnet.
By way of contrast, it will be our concern to point out the crucial disanalogies
between the two cases.

In Heisenberg's well-known model from 1928, a ferromagnet is construed
as an infinite array of spin-$\frac{1}{2}$ magnetic dipoles,
where spin-spin interactions between neighbours tend to align the dipoles.
Obviously, the model shows complete symmetry under spin rotations and
the microscopic Lagrangian is therefore SO(3)-invariant.
At high temperatures, thermal oscillation in the ferromagnet will
lead to randomized domain-like spin correlations at all length scales.
The ferromagnet therefore shows no macroscopic magnetization;
this is true at least in the absence of an external field, whereas
the ferromagnet behaves in the high temperature regime as a paramagnet.

Below a critical point, the Curie temperature, the ferromagnetic tendency
of the dipoles to align prevails over the thermal fluctuations.
We obtain a phase transition by means of an SSB of the SO(3) rotation symmetry.
In the low-temperature regime the ferromagnet will show
a spontaneous macroscopic net magnetization.


\subsection{Some General Remarks on SSB and QFT}
\label{QFT}

As suggested by the ferromagnet case,
SSB is generally characterizable as a scenario where the Lagrangian
(or equations of motion) of a physical system possesses a symmetry
that is not obeyed by the states of the system
(solutions of the equations of motion).
In particular, the energy ground state appears to be asymmetric.
This may at first seem odd since it clashes with a rather evident
principle, known as Curie's principle, stating that the asymmetries
in the effects must be found in the causes (or, conversely, that
the symmetries in the causes must be found in the effects).%
\endnote{The original source is \cite{curie1894};
for a systematic discussion of the status of Curie's principle
see for instance \cite{chalmers70}.
In a recent paper, \cite{earman2004c} addresses the principle's
connections to QFT---thereby repeating his worries about the
Higgs mechanism as already expressed in his quote in the introduction.}
But of course the actual breaking of, for instance, the dipole rotations
of the ferromagnet will in fact be caused only by an ever so small asymmetry
in the spin-spin alignments. Moreover, the new ground state of the system
after SSB has taken place shows a degeneracy such that
the system after SSB plus the total set of degenerated ground states
retains the initial symmetry of the system before SSB.

Our presentation in the two following sections \ref{toy} and \ref{gsw}
will discuss the Higgs (toy) model on the ``classical'' level of Lagrangians
only, without delving into the technicalities of a proper QFT analysis.
This may at first seem inappropriate.
For in the case of QFT the ground or vacuum state degeneracy is intimately
connected with the disturbing property of unitarily inequivalent representations
of the canonical commutation relations of the field operators---a direct
consequence of the fact that QFT systems are modelled as systems with
an infinite number of degrees of freedom.
There is therefore no unique QFT vacuum; any representation may be
associated with its own vacuum state, all of which unitarily---and,
hence, supposedly physically---inequivalent.
This then raises all the worries about Curie's principle again, and,
what is more, the particular occurrence of SSBs in QFTs seems
to be a direct consequence of the fact that the symmetries in question
cannot be unitarily implemented.

Pressing as these worries are, they will nevertheless not be
of our concern here (as should be clear from the introduction already).
The main focus of our analysis lies on the particular
\emph{gauge symmetry aspect} of the Higgs argument---and the main premise
of our argument will be that a gauge symmetry is merely conventional and
that it can therefore not be considered the source of a real physical mechanism.
This aspect can very well be brought to light on the Lagrangian level already.
Since our crucial arguments can be given from such a less complicated
point of view, we need not delve into QFT matters.

Another worry could be that even in the ferromagnet case it is necessary
to model the system in the thermodynamic limit of infinitely many dipoles
in order to obtain a phase transition---and that in this respect the analogy
with the Higgs mechanism is greater than assumed above.
While this is a technical question that deserves a rigorous
technical discussion (see, for instance, \citealt{ruetsche2006}),
the question whether, in this respect, there exists an analogy or not
does not nevertheless touch upon the clear disanalogy between the ferromagnet
and the Higgs case as far as the difference in the nature of the considered
symmetries is concerned. And it is, again, only this latter aspect
on which we will focus.

Hence, whether or not a genuine Higgs mechanism can finally be built on a rigorous
QFT approach circumventing the problem of unitarily inequivalent representations
and whether or not one day the infinity limes will conceptually be well understood,
it is in no case acceptable to claim that the breaking of a merely conventional
gauge symmetry plays a crucial role in establishing such a mechanism. Since this, however,
seems to be the case in present accounts, it is legitimate and perhaps necessary
to focus on just this single gauge symmetry aspect of the Higgs argument.


\subsection{The Higgs Mechanism in the $U(1)$ Toy Model}
\label{toy}

It is easiest to get the idea of the Higgs mechanism
by considering the $U(1)$ theory as the simplest model.
This toy model includes all the relevant features and
allows us to keep track of the argument structure more easily
than in the physically relevant model of the electroweak
$SU(2) \times U(1)$ gauge theory.
Our presentation follows the one in
\citet[Chap. 14.6--14.9]{halzen+martin84}.

We start from a Lagrangian with a complex scalar field
$\phi = -\frac{1}{\sqrt{2}} (\phi_1 + i \phi_2)$
and coupled gauge field $F^{\m\n}$:
\be
\label{higgs1}
{\cal L}' = -\frac{1}{4} F_{\m\n} F^{\m\n}
 + |(\del_\mu - iq A_\mu) \phi|^2 - \mu^2 |\phi|^2 - \l |\phi|^4.
\ee

The first two terms are kinetic, the last two describe a potential
$V(\phi) = \mu^2 \phi^* \phi + \l (\phi^* \phi)^2$.
We consider only the case $\l>0$, where the $\phi$-field is self-interacting
because of the $\phi^4$-term. From the two possibilities for the sign of $\m^2$,
the case $\mu^2>0$ simply leads to the theory of a massive scalar field
analogous to the Klein-Gordon-Lagrangian ${\cal L}_{KG} = \frac{1}{2}
(\del_\mu \f)(\del^\mu \f) - \frac{1}{2} m^2 \f^2$.
But here we are interested in the case $\mu^2<0$.
This case shows two distinctive features:
we get a negative mass term $\mu$, and $V(\phi)$ is no longer a simple parabola
but possesses energy minima with $\frac{\del V}{\del \phi}=0$ at $\phi = \pm v$,
where $v := \sqrt{\frac{-\m^2}{\l}}$. More precisely,
$V(\phi_1,\phi_2)$  now has the form of a ``Mexican hat''
with $\phi_1^2 + \phi_2^2 = v^2$ over the $(\phi_1$, $\phi_2)$-plane.
In other words, in the energetically favoured state, the ground state,
the global and local $U(1)$ symmetry is broken.

Now the first decisive step follows.
We rewrite $\phi$ as a field expansion of the vacuum state:
\be
\label{phi_expansion}
\phi = \frac{1}{\sqrt{2}} ( v + \eta + i \xi )
\ee
with real $\eta$, $\xi$.
This ansatz clearly violates $U(1)$;
after inserting (\ref{phi_expansion}) into (\ref{higgs1}) we get
\bea
\label{higgs2}
{\cal L}'' &=& -\frac{1}{4} F_{\m\n} F^{\m\n} + \frac{1}{2} q^2 v^2 A_\mu A^\mu
 + \frac{1}{2} (\del_\mu \eta)^2 + \frac{1}{2} (\del_\mu \xi)^2 \nonumber\\
&& - v^2 \l \eta^2 - q v A_\mu \del^\mu \xi + \mathcal{O}(\mbox{fields}^3).
\eea

What is the particle content of ${\cal L}''$?
Obviously we get a massive real scalar field $\eta$,
a massless $\xi$-field, and a massive vector field $A_\mu$.
The existence of a massive vector field is exactly our goal,
and the existence of the massless boson field is predicted by the
\emph{Goldstone theorem}:
For any generator of a symmetry that is broken in the ground state there exists
a massless scalar Goldstone boson. In solid state physics, for instance, such
Goldstone bosons are known as energy modes like phonons, plasmons, spin waves, etc.

Let us now compare the degrees of freedom of ${\cal L}'$ and ${\cal L}''$.
Originally, $A_\mu$ as well as $\phi$ had two degrees of freedom,
there are four physical degrees of freedom in ${\cal L}'$ altogether.
In ${\cal L}''$, however, we seem to have $1+1+3=5$ degrees of freedom.
But this is impossible, since we can hardly change the physical content
of our theory by merely transcribing it.
It turns out, indeed, that the degree of freedom of the Goldstone boson
is unphysical insofar as it can be made to disappear by a suitable gauge.
To see this we may rewrite (\ref{phi_expansion}) in polar coordinates
\be
\label{phi_polar}
\phi = \frac{1}{\sqrt{2}} (v + H ) e^{ i\frac{\q}{v} }
\ee
and choose the particular gauge
\be
\label{gauge}
A_\mu \to A_\mu - \frac{1}{qv} \del^\mu \q
\ee
in ${\cal L}'$. This is the second decisive step, since now we get
independence of $\q$; and finally
\bea
\label{higgs3}
{\cal L}''' &=& -\frac{1}{4} F_{\m\n} F^{\m\n} + \frac{1}{2} q^2 v^2 A_\mu A^\mu
   + \frac{1}{2} (\del_\mu H)^2 + \l v^2 H^2 \nonumber\\
&& - \l v H^3 - \frac{1}{4} \l H^4 + \frac{1}{2} q^2 A_\mu A^\mu H^2
   + q^2 v A_\mu A^\mu H.
\eea

From this it becomes apparent that we are indeed only dealing with a theory
of a real scalar field, the Higgs field $H$ with mass $m_H = \sqrt{2\l} v$,
and some vector field $A_\mu$ with mass $m_A = qv$.
The redundant degree of freedom of the Goldstone boson is in fact absorbed
into the longitudinal polarization of the gauge boson. It is precisely
this transcription of degrees of freedom---because of the non-invariance
of the ground state---which is usually called the ``Higgs mechanism''.


\subsection{The Higgs Mechanism in the Electroweak Model}
\label{gsw}

The more complex model of the GSW electroweak theory
considers a Higgs mechanism which starts in the Goldstone mode
from an $SU(2)$ doublet $\phi$
with $2 \cdot 2 = 4$ degrees of freedom
and a gauge coupling
to three massless vector bosons $\mathbf{W}_\m$
corresponding to the generators of SU(2)
and one massless vector boson  $B_\m$
corresponding to the generator  of  U(1),
which means $4 \cdot 2 = 8$ degrees of freedom
for the vector fields and twelve degrees of freedom in toto.

In the Higgs mode we go over to
three massive vector bosons $W_+$, $W_-$, $Z_o$,
that is $3 \cdot 3 = 9$, one massless photon $\g$ with two
and one massive Higgs scalar $H$ with one degree of freedom,
altogether again twelve degrees of freedom.

In the full-blown model, the Higgs mechanism is also used
to generate the lepton masses.
It is, however, not necessary for the purposes
of this paper to delve into the details of this application,
since it is really the same logic applied to another case.


\section{A Threefold Analysis}
\label{threefold}

We start our threefold analysis (as regards the ontological, explanatory and
heuristic value of the Higgs model) with three observations.
Needless to say, all these observations apply both to the U(1) toy
as well as to the electroweak model.

First observation: all three considered Lagrangians are mathematically equivalent
in the sense that they belong to a different choice of variables and gauge
\be
{\cal L}' \sim {\cal L}'' \sim {\cal L}'''.
\ee
More precisely, from ${\cal L}'$ we go over to ${\cal L}''$
by rewriting the field variable $\phi$ in terms of $v$, $\eta$ and $\xi$
according to (\ref{phi_expansion}), which does, however,
lead to a spurious degree of freedom.
This gauge freedom is removed by the transcription
(\ref{phi_polar}) instead of (\ref{phi_expansion}) for $\phi$,
now written in terms of $v$, $H$ and $\q$,
together with the particular gauge fixing (\ref{gauge}),
by which the transition from ${\cal L}'$ to ${\cal L}'''$ is accomplished.
From this observation the suspicion immediately arises that the whole
``mechanism'' consists in a mere shuffling of degrees of freedom!

The second observation is that all three Lagrangians are in fact invariant
under U(1) and that the symmetry is ``broken'' not on the level of the Lagrangians,
but only on the level of their ground states.

The third observation is that indeed the first Lagrangian ${\cal L}'$
with parameter choice $\l>0$, $\mu^2<0$ does not allow for any quick,
literal interpretation, since here we are facing the obscure case of
a $\phi$-field with imaginary mass $\mu$.

The second observation already demonstrates the reason why some authors
prefer the terminology of ``hidden symmetry'' instead of ``SSB''
(e.g.\ \citealt{oraifeartaigh79}). Given our remarks in Sec.~\ref{QFT}
that the system after ``SSB'' plus the set of ground states
retains the initial symmetry of the system before ``SSB'',
this is certainly conceptually far more precise.
The third observation may perhaps be questioned
from a rigorous QFT point of view (see below),
and so it will be in particular the first observation together with
the conventional status of gauge symmetries
which eventually undermines the prospects of an ontological picture
of the Higgs mechanism.


\subsection{Ontology of the Higgs Mechanism?}
\label{ontol}

As already mentioned in Sec.~\ref{ferromagnet}, it is a widespread
view that the SSB of the Higgs mechanism is of the same kind
as in the case of spontaneous magnetization in the ferromagnet.
One is then tempted to regard the Higgs mechanism as a dynamic evolution,
a real process in time with a dependence on temperature
(e.g.\ \citealt{linde79}; cf.\ also \citealt[633]{huggett2000},
and \citealt[Sec.~4]{balashov2002}).
The following passage from a panel discussion of the 1996 Boston
University Conference on the Conceptual Foundations of Quantum Field Theory
highlights this:
\bq
\bd
\item[]Huggett: What is the mechanism, the dynamics for spontaneous
                symmetry breaking supposed to be? ...
                I mean isn't this a dynamic evolution, something
                that happens in the history of the universe?
\item[]Coleman: Oh, it happens with temperature, yeah.
                Typically at high temperature you're very far from the ground state
                but the density matrix or whatever has the symmetry.
                Have I got it right, Steve? You were one of the first to work this out.
\item[]Weinberg: Yeah, it doesn't always happen, but it usually happens.
\item[]Coleman:  Yes, typically at high temperatures the density matrix
                has a symmetry which then disappears as the temperature gets lower.
                But that's also true for ordinary material objects.
                ... it's the same thing.
                The difference between the vacuum and every other
                quantum mechanical system is that it's bigger.
                And that's from this viewpoint the only difference.
                If you understand what happens to a ferromagnet
                when you heat it up above the Curie temperature,
                you're a long way towards understanding one of the possible ways
                it can happen to the vacuum state. \cite[Chap. 26]{cao99}
\ed
\eq

Without doubt, the ferromagnet's SSB allows a straightforward
realistic interpretation: the observable change of the macroscopic net magnetization.
In this case we \emph{do} have a real dynamic process in time with a phase transition
at Curie temperature and with a real instantiation of the underlying symmetry:
the rotational degrees of freedom of the elementary magnetic dipoles of the ferromagnet.%
\endnote{It is not necessary to delve into any sophisticated philosophical debate
about realism here. We simply use a minimal and, perhaps, commonsensical notion
of physical reality, where physical quantities are considered to be connected with
observable consequences---and we take this notion, for the purpose of this discussion,
as an unproblematic notion.}
But nothing like that holds in the case of the Higgs mechanism's SSB,
as the following three objections show.

First, the transition from the Goldstone mode to the Higgs mode cannot be
understood as a real process in the world, because of our third observation:
the impossibility of a realistically interpretable particle content of ${\cal L}'$.
A typical handwaving argument at this point could be that we may nevertheless
consider the Goldstone regime portrayed by ${\cal L}'$ as real, since
the $\phi^4$-term is dominating the imaginary mass term of order $\phi^2$,
which means that the latter may simply be neglected at the high energies
prevalent in the early cosmos.
While such an argument is at best satisfying from a pragmatic and instrumentalist
perspective, it still leaves open the ontological question of an appropriate
interpretation of physical entities with imaginary masses.%
\endnote{One might, perhaps, speculate about the introduction of
new physical principles here---for instance a new variant of the
``Cosmic Censorship'', where Nature forever hides imaginary masses from our eyes.
But nothing like that has been worked out by anyone yet.}

One has to admit, however, that our way of presenting this objection
is in a sense based on an all too naive view of interpreting Lagrangians.
From a more rigorous quantum field theoretic perspective the definition of mass
depends on the definition of the ground state of the theory---and it is the
whole point about the Higgs mechanism that ${\cal L}'$ doesn't give us the
``true'' ground state. But this only underlines our overall suspicion of
${\cal L}'$ from the more elaborate viewpoint of QFT.

A second objection concerns the reality of the Goldstone bosons
(recall Weinberg: ``Goldstone bosons ... are in some sense real'').
In the case of the ferromagnet the Goldstone bosons indeed exist
as long-range spin oscillations, but the application of the Goldstone theorem
in the case of a \emph{gauge symmetry} leads to spurious, unphysical degrees
of freedom, which can be transformed away by our conventional choice of gauge.
There simply seems to be no sense in which Goldstone bosons can be given
a realistic interpretation (recall again Earman:
``... a genuine property like mass cannot be
gained by eating descriptive fluff, which is just what gauge is'').

This point is directly connected with the third, most important objection
against the analogy between the ferromagnet and the Higgs case,
and thereby against any ontological picture of the Higgs mechanism.
Whereas in the case of the ferromagnet SO(3) is instantiated by real rotations
of the dipoles, \emph{quantum gauge transformations possess no such real instantiations.}
This was already highlighted in the introduction:
neither global nor local unitary gauge transformations are observable,
the status of gauge symmetries is a non-empirical and merely conventional one.

A few explanations are in order here.
The conventional nature of the choice of gauge should be clear.
Also, in view of global quantum gauge transformations, the claim about
their non-observable or non-empirical status is certainly uncontroversial,
since it is well-known that the expectation value
$\langle \hat A \rangle = \frac{\langle \psi | \hat A | \psi \rangle }{\langle \psi | \psi \rangle}$
of a quantum observable $\hat A$ is invariant under $\psi \to \psi e^{i \chi}$.
Perhaps this is not immediately clear for local quantum gauge transformations
$\hat U(x)=e^{i \chi(x)}$. For here one might argue that, for instance,
the eigenvalue $p$ of the momentum operator $\hat{p}_\mu = - i \partial_\mu$
for a plane wave $\psi=e^{ipx}$ changes for a locally phase transformed wave
$\psi' = \hat{U} \psi$ into $p + \partial_\mu \chi(x)$---and that
this is a physically significant change.
To see the fallacy of this argument, consider the wave function
$\Psi(x) = \langle x | \phi \rangle$ in the position representation $\ket{x}$,
where the $\big\{\ket{\phi}\big\}$ span an abstract Hilbert space.
Here, one immediately sees that a proper understanding of local gauge
transformations $\ket{x'} = \hat U \ket{x}$ is in terms of changes in $\ket{x}$,
i.e.\ merely conventional changes in the position representation.
And, of course, such changes affect all Hilbert space operators
$\hat O' = \hat{U} \hat O \hat{U^+}$ as well.
In the above example we must therefore use the appropriate covariant
momentum operator $\hat{p}'_\mu = \hat{U} \hat{p}_\mu \hat{U^+}$,
which in application to $\psi'$ leads again to eigenvalue $p$.
This demonstrates the non-empirical nature of local gauge transformations
(cf.\ \citealt{healey2001}, \citealt{brading+brown2004}, \citealt{lyre2004b}).

On the positive side, to characterize a theory as a gauge theory
with gauge group $G$ means to single out the form of the field
strength interaction tensor, defined in fibre bundle terms
as the curvature tensor of the appropriate $G$-connection.
Here of course the analogy between gauge transformations of a particular
group $G$ and coordinate transformations ends: while it must in principle
be possible to give a coordinate covariant formulation for \emph{any} physical
theory (rendering the principle of general covariance in one specific sense
physically insignificant), the claim that a certain interaction field exists
in nature and that, as such, a particular gauge group $G$ applies
is of course physically significant. Nevertheless, by analogy with
coordinate transformations, the $G$-transformations themselves conform
to nothing more than conventional changes of a fibre bundle representation
and do not possess any real instantiations.%
\endnote{\label{gaugeprin}From all the above the logic of the gauge
principle should also become clear: the demand of local gauge invariance prompts
the introduction of a covariant derivative $D_\mu = \hat{U} \partial_\mu \hat{U^+}$.
In the usual textbook presentations (e.g.\ \citealt[316]{halzen+martin84}),
however, the gradient of the phase is written in terms of a vector field,
where also the dimensions of a charge come in:
$\partial_\mu \chi(x) = - q A_\mu(x)$. This suggests a reading of
the covariant derivative $D_\mu =  \partial_\mu + i q A_\mu(x)$ as if the
existence of a gauge potential $A_\mu$ were enforced. But one must
be aware of the fact that a thus introduced potential is, in fibre bundle
terminology, a \emph{flat connection} only.
That is to say the physically significant curvature tensor, the derivative
of the connection, is still zero. Whether, in fact, a particular curvature
or gauge field interaction tensor is non-zero and is as such realized by nature,
is of course an \emph{empirical input} and cannot be dictated by demanding
local gauge invariance (see footnote \ref{recipe} and references therein).}

To sum up: the three negative results---no real instantiations of imaginary masses,
no real instantiations of Goldstone bosons, and
no real instantiations of gauge transformations---provide a clear
answer to our overall question:
no ontological picture of the Higgs mechanism seems tenable;
the possibility of an as yet undiscovered process or a mechanism supplemented
to the exposition given in sections \ref{toy} and \ref{gsw} notwithstanding.
But as far as the exposition in \ref{toy} and \ref{gsw} is concerned,
the Higgs mechanism ``does not exist''.


\subsection{Explanatory Value of the Higgs Mechanism?}

Once an ontic interpretation of the Higgs mechanism is blocked,
it seems natural to ask further for the epistemic and explanatory value
of the hypothesis.
After all, the Higgs mechanism was introduced to explain the masses
of the elementary particles. Is such a goal reached?

We are obviously facing a rather misleading terminology again.
It is, at first, well known that the Higgs model does not allow one to predict
the individual values of the particle masses (but only certain ranges
because of general boundary conditions such as the applicability
of renormalization procedures and the like). But also a more general
explanation of masses, let alone of the nature of mass, is hardly given.
On the contrary: our ontological analysis has clearly shown that the masses
are not ``dynamically generated'' in any literal or realistic sense---the
particular values are rather put in by hand as free parameters
of the GSW Lagrangians ${\cal L}'_{GSW}$ to ${\cal L}'''_{GSW}$.

But let us consider the number of parameters of the GSW model.
In the Goldstone mode we count the two
coupling constants $g$ and $g'$ of $SU(2) \times U(1)$ and
two further parameters $\m$ and $\l$ of the Higgs potential,
i.e.\ four parameters in total.
In the Higgs mode we have again the two coupling constants
$g$ and $g'$ together with three masses:
$m_H$ of the Higgs as well as $M_W$ and $M_Z$ of the weak bosons.
These five parameters, however, depend in a particular manner
on the Weinberg angle $\theta_W$:
\be
\label{weinberg1}
e = g \sin \theta_W = g' \cos \theta_W
\ee
and
\be
\label{weinberg2}
M_W = M_Z \cos \theta_W .
\ee
From these relations it becomes clear that we are effectively dealing
with only four parameters, as one would expect from the Goldstone mode,
yet the dependency relation (\ref{weinberg2}) is frequently considered
a \emph{prediction} of the model---and this looks like a considerable
explanatory strength of the Higgs mechanism.

But here again objections have to be raised.
Already from the empirically known V-A structure of the electroweak current
the relation (\ref{weinberg1}) is required as well as the particular mixing
of states, in which the neutral gauge boson fields $A_\mu$ and $Z_\mu$
must be written. The mixing of $g$ and $g'$ (i.e.\ $\theta_W \neq 0$)
leads to a theory in which $SU(2)$ and $U(1)$ are linked in a non-trivial
manner, the requirement $\theta_W \neq 0$ then induces
(\ref{weinberg1}) and (\ref{weinberg2}).
The dependency relations are in this way directly built in
to the mathematical structure of $\mathcal{L}'''_{GSW}$.
The final upshot of the whole discovery history of GSW, however,
was that under the condition of constructing a renormalizable theory,
it was mandatory to have the non-trivial symmetry mixing and, hence,
to get dependencies in the form of (\ref{weinberg1}) and (\ref{weinberg2}).
This fact shows the supposed explanatory value of GSW in a new light,
since even if one were to construe GSW from scratch one could,
under the condition of renormalizability, derive $\mathcal{L}'''_{GSW}$
directly without the detour through the Higgs mechanism.
The empirical validity of the relations (\ref{weinberg1}) and (\ref{weinberg2})
is after all only an indicator of the empirical appropriateness of the Higgs mode
Lagrangian---nothing less, but certainly also nothing more.


\subsection{Heuristic Value of the Higgs Mechanism}

We seem to be left with a disastrous result:
nothing is explained by the Higgs mechanism at all!
Within the context of justification considered so far, we were not able
to single out any convincing argument in favour of the conceptual idea
of the Higgs mechanism. But certainly---given the predominance of the
Higgs story in the literature---\emph{there must be something to it.}
What could this something be? The answer lies in the context of discovery.

As already outlined in the introduction, the Higgs mechanism was obviously
an important heuristic tool for reconciling gauge symmetry and renormalizability
in the 1960's (recall Glashow: ``I completely missed the boat...'').
Moreover, from a heuristic perspective every physicist will immediately support
the view that ${\cal L}'_{GSW}$, the Lagrangian from which the GSW-Higgs model
takes its starting-point, has a far ``simpler'' mathematical structure
than ${\cal L}'''_{GSW}$. That is to say, it is almost impossible to invent
or to discover ${\cal L}'''_{GSW}$ from scratch.
It is, instead, more than convenient to have some ``guiding story'' leading
from ${\cal L}'_{GSW}$ to ${\cal L}'''_{GSW}$.
And this is all the more true insofar as $\mathcal{L}'''_{GSW}$
seems to describe an essential trait of reality.

The Higgs mechanism as the guiding story leading from
${\cal L}'_{GSW}$ to ${\cal L}'''_{GSW}$ within the context of
the early discovery of GSW certainly had its overwhelming heuristic value,
as the quotations in the introduction show.
And it is also true that, from a purely mathematical point of view, ${\cal L}'_{GSW}$
is written down in a far more tractable representation than ${\cal L}'''_{GSW}$.
But at the end of the day this is only a matter of mathematical representation.%
\endnote{\label{foot}
A further note of clarification: the reader might perhaps be puzzled
by our claims about the equivalence of the three Lagrangians on the one hand
and the impossibility of a direct realistic interpretation of ${\cal L}'$
as opposed to ${\cal L}'''$ on the other hand. There seems to be a tension
between our first and third observation in the beginning of section \ref{threefold}.
And indeed, observation three hinges on stressing a ``quick and literal''
interpretation of the particle content of a Lagrangian by simply looking
at the mass terms. On the basis of our analysis we may now of course say
that such a quick and literal interpretation of ${\cal L}'$ cannot directly be gained,
but is rather indirectly revealed by the direct interpretation of ${\cal L}'''$.
}
From the physical point of view and given the devastating analysis
in the last two sections we could---or actually should---introduce
the GSW model by writing down ${\cal L}'''_{GSW}$ directly.

The upshot is that the Higgs mechanism had its heuristic value
only within the context of discovery, whereas within the context of justification
this very ``mechanism'' should rather be considered a kind of
\emph{Wittgensteinian ladder}: once ${\cal L}'''_{GSW}$ is introduced
we may without further ado forget about its heuristic derivation.


\section{Conclusion}

We have clearly seen that what the Higgs mechanism is all about is a mere
reshuffling of degrees of freedom. It certainly does not describe
any dynamical process in the world---no ontic interpretation of the Higgs mechanism
is tenable, since neither imaginary mass particles, nor GSW Goldstone bosons,
nor quantum gauge transformations have any real instantiations in nature.
The whole story about the ``mechanism'' is just a story about ways of
representing the theory and fixing the gauge.
We have also seen that no concrete physically explanatory value
of the Higgs mechanism within the context of justification can be pointed out,
but rather a heuristic value within the early context of discovery of GSW.

Needless to say, on the other hand, our exclusively conceptual analysis
does not involve any direct arguments against the possible existence of the
as yet undetected Higgs boson: ${\cal L}'''_{GSW}$ may very well describe reality,
after all this is a purely empirical question.
In fact, given the strong empirical evidence for GSW, we must definitely assume
that the Higgs boson exists. It intimately belongs to the structure and predicted
particle content of the theory, as most clearly revealed by an interpretation of
the kinetic terms and mass terms in ${\cal L}'''_{GSW}$ (bearing in mind footnote
\ref{foot}). And this is what GSW and its empirical evidence commits us to.
But GSW does not commit us to a story, called ``Higgs mechanism'', that pretends
to deliver a dynamical picture about how the predicted particles come into being.

Perhaps our critical analysis may have an impact on the power of persuasion
of ${\cal L}'''_{GSW}$ regarding a deeper understanding of the nature of mass.
It has often been pointed out that the Standard Model has too many
free parameters, including all the mass values of the elementary particles.
Our analysis certainly makes the seemingly ad hoc character of the GSW model
even more transparent. Where, if not from ${\cal L}'_{GSW}$ and the supposed
spontaneous breaking of a gauge symmetry, does the \emph{structure}
of ${\cal L}'''_{GSW}$ arise from?

We will learn---should the Higgs boson one day be found---that
$\mathcal{L}'''_{GSW}$ indeed describes an essential trait of reality.
In the light of our critical analysis of the Higgs mechanism,
this raises a series of fundamental questions.
Why does the Higgs field occur in nature together---in one package, so to speak---with
the other particles of the GSW theory? What, if at all, is the true connection
between the Higgs field and the masses of the particles, if it is not a story
about the breaking of a gauge symmetry of the ground state of ${\cal L}'_{GSW}$?
These questions are undoubtedly fascinating questions, but also undoubtedly
questions about the physics beyond the Standard Model and, as such,
an open task to physics, not to philosophy.


\section*{Acknowledgements}

I am deeply grateful to Tim Mexner-Eynck especially for helping me
with some technical details of the present paper, but, what is more,
for years of a unique collaboration between philosophy of science
and theoretical physics.
Many thanks also to Gernot M{\"u}nster, Wolfgang Unger
and two anonymous referees for useful remarks.

\begingroup

\theendnotes

\endgroup


\newpage

\bibliographystyle{apalike}


\end{document}